\documentclass[draft]{agujournal2019}
\usepackage{url} 
\usepackage{lineno}
\usepackage[inline]{trackchanges} 
\usepackage{soul}
\usepackage{natbib}
\usepackage{amsmath, amssymb}
\setcitestyle{authoryear,open={(},close={)}} \usepackage{hyperref}
\usepackage{appendix}
\usepackage{lmodern}
\usepackage{textgreek}
\draftfalse

\journalname{JGR: Machine Learning and Computation}

\begin{document}

%
%


\title{Accelerating the discovery of steady-states of planetary interior dynamics with machine learning}

%
%




\authors{Siddhant Agarwal\affil{1}, 
        Nicola Tosi\affil{1},
        Christian H{\"u}ttig\affil{1},
        David S. Greenberg\affil{2},
        Ali Can Bekar\affil{2}
}

 \affiliation{1}{Institute of Planetary Research, German Aerospace Center (DLR), Berlin}
 \affiliation{2}{Model-driven Machine Learning, Helmholtz-Zentrum Hereon, Geesthacht}




\correspondingauthor{Siddhant Agarwal}{siddhant.agarwal@dlr.de}



\begin{keypoints}
\item Achieving a statistical steady-state in complex mantle convection simulations can be computationally challenging.
\item Optimal initial conditions predicted by neural networks allow reaching a steady-state 3.75 times faster than standard initial conditions.
\item This machine learning model can be easily shared and used to accelerate mantle convection research, such as for deriving scaling laws.
\end{keypoints}

%
%

%
%


\begin{abstract}

Simulating mantle convection often requires reaching a computationally expensive steady-state, crucial for deriving scaling laws for thermal and dynamical flow properties and benchmarking numerical solutions. The strong temperature dependence of the rheology of mantle rocks causes viscosity variations of several orders of magnitude, leading to a slow-evolving ``stagnant lid'' where heat conduction dominates, overlying a rapidly-evolving and strongly convecting region. Time-stepping methods, while effective for fluids with constant viscosity, are hindered by the Courant criterion, which restricts the time step based on the system's maximum velocity and grid size. Consequently, achieving steady-state requires a large number of time steps due to the disparate time scales governing the stagnant and convecting regions.

We present a concept for accelerating mantle convection simulations using machine learning. We generate a dataset of 128 two-dimensional simulations with mixed basal and internal heating, and pressure- and temperature-dependent viscosity. We train a feedforward neural network on 97 simulations to predict steady-state temperature profiles. These can then be used to initialize numerical time stepping methods for different simulation parameters. Compared to typical initializations, the number of time steps required to reach steady-state is reduced by a median factor of $3.75$. The benefit of this method lies in requiring very few simulations to train on, providing a solution with no prediction error as we initialize a numerical method, and posing minimal computational overhead at inference time. We demonstrate the effectiveness of our approach and discuss the potential implications for accelerated simulations for advancing mantle convection research.

\end{abstract}

\section*{Plain Language Summary}

Simulating the dynamics of the rocky interiors of planets can be particularly demanding when seeking a statistical steady-state. Yet, such simulations play a key role in planetary research for instance in benchmarking numerical codes and deriving parameterizations for convective heat transfer and mixing. We show that a neural network trained on fewer than $100$ simulations can accurately predict the horizontally-averaged temperature profile of the final steady-state. If these profiles are used as initial conditions for new numerical simulations, the corresponding steady-state for the simulation in two dimensions and all the solution variables is attained nearly four times faster than when using other typical initializations. By speeding up thermal convection simulations, machine learning promises to enhance our understanding of the dynamics of planetary interiors.

%
%

\section{Introduction}

\subsection{Challenges in achieving steady-state in mantle convection simulations}

Simulating natural convection phenomena, particularly in mantle dynamics, often necessitates achieving a steady-state, a condition where parameters of interest either remain constant or oscillate around a mean value. This equilibrium phase is essential for deriving scaling laws for convective heat transfer \citep[e.g.,][to name just a few studies spread over three decades of research]{christensen1984,solomatov1995,grasset1998,korenaga2010,vilella2018,ferrick2023}, validating numerical codes \citep[e.g.,][]{blankenbach1989,king2010,tosi2015}, studying chaotic mixing processes \citep[e.g.,][]{farnetani2003,coltice2006,samuel2012,vankeken2014}, and determining characteristic flow wavelengths \citep[e.g.,][]{grigne2005,hoink2010,phillips2010}. 

In computational fluid dynamics (CFD), simulations typically overcome non-linear terms such as advection by advancing the temperature over discrete time intervals. The time-step size is constrained by the maximum velocity within the system and by the grid size as per the Courant criterion, which is crucial for maintaining numerical stability. 
This approach is effective for fluids with constant viscosity, where steady-state conditions are rapidly achieved. In mantle convection, however, this criterion becomes problematic due to the extensive viscosity range (often spanning 10 orders of magnitude). This variability leads to cold upper regions that remain static forming a so-called stagnant lid — a ``frozen'' layer where heat transfer occurs very slowly, primarily through conduction, while deeper and hotter regions exhibit vigorous convection \citep[e.g.,][]{moresi1995}. Below this layer, as the temperature increases, convection dominates, providing a heat flux into the stagnant lid that affects its time evolution \citep[e.g.,][]{breuer2015}. The coexistence of these two distinct thermal regimes implies that the temperature distribution of such a system evolves over vastly different time scales. This often requires several hundred thousand to a few million time iterations before the system reaches a steady-state. Any reduction in the number of iterations needed to reach the final state can thus offer significant computational and time savings for mantle convection studies.  

\subsection{Machine learning in fluid dynamics}

The high computational cost of running fluid dynamics simulations has made them a prime area of research in scientific machine learning \citep[e.g.,][]{BruntonMLReview, LinoMLReview}. This is not only true for surrogate modeling strategies where trained machine learning models serve to approximate the forward problem, but also for hybrid strategies, where machine learning is used to accelerate partial differential equations solvers without compromising their numerical accuracy \citep[e.g.,][]{tompson2017,Kochkov,Ekhi}.

Likewise, machine learning remains a promising avenue for accelerating discovery in geo- and planetary physics . Recent successes include modeling of seismic wave propagation \citep{biotsquirt}, seismic fault segmentation \citep{wu_faultseg3d_2019}, wave inversion \citep{rasht-behesht_physics-informed_2022}, predictions of pairwise planetary collisions from Smooth Particle Hydrodynamics simulations \citep{nbodysph}, trace gases prediction \citep{azmi}, weather prediction \citep[e.g.,][]{graphcastnet, fourcastnet}, air pollution forecasting based on a foundation model \citep{aurora}, probabilistic characterization of the interior structure of planets \citep{exomdn}, 
prediction of surface heat flow from seismic structure for modeling ice sheet dynamics \citep{zhang_ritzwoller_2024}, pressure and temperature predictions for mineral combinations in the Earth's lithosphere based on experiments \citep{qin2023}, and gravitational mass and moments modeling of Jupiter \citep{neuralcms}. The recent publication of a few review articles on data-driven methods in geophysics and geodynamics \citep[e.g.,][]{bergen2019review, Morra2020, review_geophysics} underscores the field's increasing significance.

Closer yet to the topic of convection simulations, \citet{Shahnas2020} used a feedforward neural network (NN) to predict the surface heat flux and mean temperature of steady-state simulations as a function of parameters such as the Rayleigh number ($Ra$), the core-to-planet radius ratio, and the presence of melting. \citet{agarwal2021b} demonstrated that two-dimensional surrogate models of thermal evolution of planetary interiors could be modeled using autoencoders for dimensionality reduction and then by using Long Short-Term Memory (LSTM) networks for advancing the compressed representations of this flow in time. This was a follow-up study to \citet{agarwal2020}, where an NN was used to predict the horizontally-averaged temperature profiles from thermal evolution simulations. \citet{akbari} leveraged Proper Orthogonal Decomposition and LSTM networks to build reduced-order models for data assimilation in Rayleigh–B\'{e}nard convection. \citet{Boroumand} used deep convolutional networks to estimate the Prandtl number and $Ra$ from two-dimensional snapshots of convection simulations. \citet{xgboost} used eXtreme Gradient Boosting to predict the Nusselt number as a function of $Ra$ and aspect ratios of the computational domain for turbulent thermal convection. The ability of machine learning methods to model high-dimensional function and parameter spaces has thus triggered tremendous interest in several areas of fluid dynamics.

\subsection{Using neural networks to predict statistical steady-state}

In this paper, we exploit the findings of \citet{agarwal2020} that an NN can predict horizontally-averaged temperature profiles as a function of simulation parameters. Here, we focus on predicting temperature profiles of statistically steady-state mantle convection simulations instead of thermal evolution simulations. We demonstrate that these predicted profiles serve as excellent initial conditions for a traditional numerical solver. By using the profiles predicted by our NN, we can determine the statistical steady-state of a given system in a fraction of the time than would be needed when starting from a typical initial condition such as an arbitrary high or low temperature, or a conductive profile. 

While the predictions of the temperature profile from the NN are expected to be relatively accurate, feeding these to numerical solvers allows obtaining other valuable information on the system, namely, (1) two-dimensional temperature, velocity, and pressure fields, and thereby the characteristic flow pattern and wavelength, as well as (2) numerically-accurate heat fluxes at the top and bottom of the domain, which can then be used to derive suitable scaling laws for convective heat transfer (see Sec. \ref{subsec-scalinglaw}). 
Hence, we propose a hybrid modeling approach, where NNs serve as an accelerator for our specific problem of predicting the statistical steady-state of thermal convection simulations. To the best our knowledge, this is the first time such an approach has been proposed in the context of mantle convection. The data used in this study as well as a trained NN are made available on HuggingFace. The trained model can be run to generate and download temperature profiles: \url{https://huggingface.co/spaces/agsiddhant/steadystate-mantle}. No installation is needed.

The outline of this paper is as follows. We first explain the methods used, including the mantle convection equations solved and the numerical setup of the simulations (Sec. \ref{subsec-pde}). We provide details on the simulation dataset (Sec. \ref{subsec-params}), on the neural network used (Sec. \ref{subsec-nn}) as well as on the other regression methods used as baselines (Sec. \ref{subsec-baseline}) and on the calculation of the stagnant lid thickness and of some basic scaling laws (Sec. \ref{subsec-lid}). In Sec. \ref{subsec-prediction}, we examine the accuracy of the NN predictions and compare it to the other regression baselines. In Sec. \ref{subsec-speedup} we compare the number of iterations needed to reach a statistical steady-state for three qualitatively different simulations that are started from different initial conditions. In Sec. \ref{subsec-scalinglaw}, we derive some basic scaling laws using simulations accelerated by NN predictions as an example application. We then conclude the paper discussing the significance of this approach and possible improvements (Sec. \ref{sec-discussion}) and summarizing our main findings (Sec. \ref{sec-conclusions}).

\section{Methods}
\label{sec-methods}

\subsection{Convection model}
\label{subsec-pde}

We solve the conservation equations of mass, linear momentum and thermal energy for an incompressible fluid with variable-viscosity and negligible inertia heated from below and from within. In non-dimensional form, these read respectively:
\begin{linenomath*}
\begin{align}
  & \nabla \cdot \boldsymbol{u} = 0, \label{eq:mass}\\
  & - \nabla p + \nabla \cdot \left( \eta \left( \nabla \boldsymbol{u} + (\nabla \boldsymbol{u})^{\textrm{T}} \right) \right) = \textrm{Ra} T \boldsymbol{e}_y, \label{eq:momentum}\\
  & \frac{\partial T}{\partial t} + \boldsymbol{u} \cdot \nabla T = \nabla^2 T + Q \label{eq:energy},
\end{align}
\end{linenomath*}
where $\boldsymbol{u}$ is the velocity vector, $p$ is the dynamic pressure, $\eta$ is the dynamic viscosity, which depends on temperature and depth (see eq. \eqref{eq:eta}), $T$ is the temperature, $\boldsymbol{e}_y$ is the unit vector in the vertical direction,  $Q$ is the internal heating rate, and $Ra$ is the reference Rayleigh number defined as 
\begin{linenomath*}
\begin{equation}
    Ra=\frac{\rho^2 c_p g \alpha \Delta T D^3}{\eta_0 k},
    \label{eq:Ra}
\end{equation}
\end{linenomath*}
where $\rho$ is the density, $c_p$ is the heat capacity, $\alpha$ is the coefficient of thermal expansion, $\Delta T$ is the temperature scale, $D$ is the domain thickness, $k$ is the thermal conductivity, and $\eta_0$ is a reference  viscosity. 

All quantities appearing in eq. \eqref{eq:Ra} are constant (i.e. their non-dimensional reference value is 1). The (non-dimensional) viscosity in eq. \eqref{eq:momentum} depends on temperature and depth according to the following rheological law:
\begin{linenomath*}
\begin{equation}
\eta(T,y) = \exp\left( -\log(\gamma) T  + \log(\beta) (1-y) \right),
\label{eq:eta}
\end{equation}
\end{linenomath*}
where $y$ is the height (0 at the bottom of the domain and 1 at the top), and $\gamma$ and $\beta$ are the maximum viscosity contrasts due to temperature and depth, respectively. With the above definition, the reference Rayleigh number \eqref{eq:Ra} is equal to 1 at the top of domain (where $T=0$ and $y=1$). By using large values of $\gamma$, we obtain a stagnant lid overlying a convecting layer with a highly super-critical Rayleigh number (see Sec. \ref{subsec-params}). 

We use the finite volume code GAIA \citep{huettig2013} to solve eqs. \eqref{eq:mass}, \eqref{eq:momentum} and \eqref{eq:energy} in a 2D rectangular domain with an aspect ratio of 4. We use a regular grid consisting of 126 and 504 grid points in the vertical and horizontal directions, respectively. We treat all boundaries as impermeable, i.e. with zero normal velocity, and free-slip, i.e. with zero shear stress. The top and bottom boundaries are isothermal (with $T(y=0)=1$ and $T(y=1)=0$) and the sidewalls insulating, i.e. with zero heat flux. 

\subsection{Parameters, dataset, and output quantities}
\label{subsec-params}

We generate a dataset of 128 simulations of steady or statistically-steady convection depending on the three parameters $\gamma$, $\beta$ and $Q$. We vary $\gamma$ between $10^6$ and $10^{10}$, $\beta$ between $1$ and $100$, and $Q$ between $1$ and $10$. This choice results in effective Rayleigh numbers ranging from $10^4$ (for $\gamma=10^6$ and $\beta=100$) to $10^{10}$ (for $\gamma=10^{10}$ and $\beta=1$).

\begin{figure}
\centerline{\includegraphics[width=1.4\textwidth]{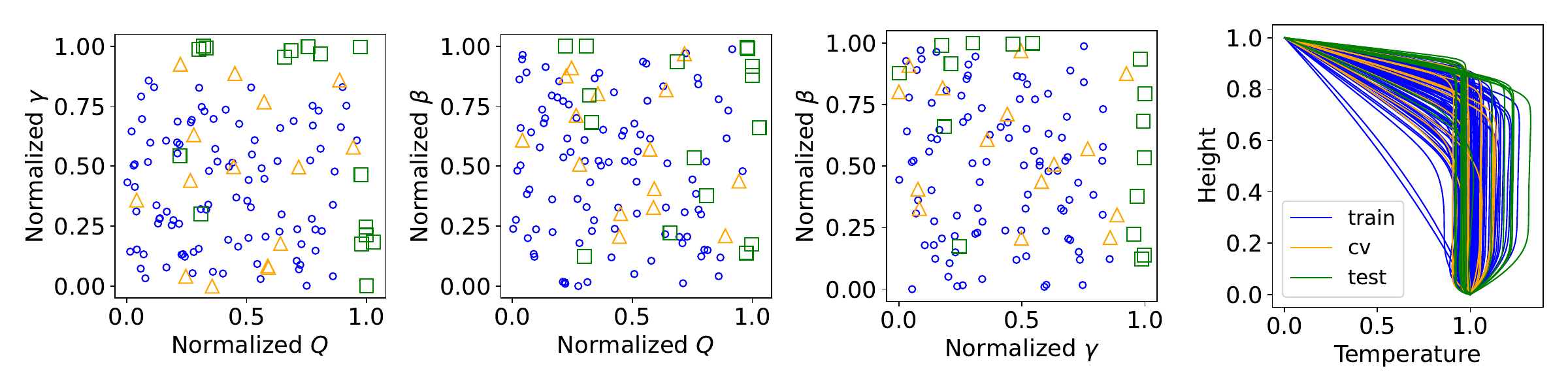}}
\caption{A visualization of the normalized simulation parameters and temperature profiles for train (blue), validation (cv, orange) and test (green) sets. The test set is deliberately picked in this manner to check for NNs ability to extrapolate slightly.}
\label{fig-dataset}
\end{figure}

Due to choice of rectangular geometry, the combination of high internal heating, depth dependence of the viscosity and fixed-temperature bottom boundary conditions can lead to internal temperatures exceeding the bottom temperature, which is set to the non-dimensional value of 1 (Fig. \ref{fig-dataset}, last column). Indeed, plane-layer convection models may need to introduce artificial internal cooling in order to reproduce temperatures obtained in a spherical shell geometry \citep{ofarrell2010,ofarrell2013}.
Also, in a more realistic mantle convection setting, the bottom temperature (i.e. the temperature of the core) would quickly adapt according to the local heat flux, at least as long as the core is liquid and can be approximated as a convecting and homogeneous body of given density and heat capacity \citep[e.g.][]{stevenson1983}. In this situation, the core would tend to heat up and the heat flux out of it to vanish as in a purely internally-heated system. Yet, for the purpose of this work, these cases provide useful tests to probe the ability of our ML predictions to reproduce non-trivial temperature profiles. 

We split the dataset of 128 simulations as follows: 97 simulations are used for training the network, 15 are used as validation (cv) to test different architectures, and 16 are kept as the test set. While training and cv simulations are chosen randomly from the same uniform distribution of the parameters, we intentionally chose our test set so that at least one of the simulation parameters is slightly outside the range of the training and cv datasets: if $Q > 9.5$ or if $\gamma > 5 \times 10^9$ or if $\beta > 95$. This allows us to test the ability of the network to extrapolate, albeit slightly given the limited amount of simulations at hand. The simulation parameters as well as the temperature profiles are visualized in Fig. \ref{fig-dataset}. As somewhat expected, the temperature profiles of some of the simulations exceed the adiabatic temperature and the heat fluxes of the profiles of the training and cv sets. 

To facilitate optimization of the NNs and other regression algorithms, we scale the parameters by their respective minimum and maximum values. For $\gamma$ and $\beta$, we take the $\log_{10}$ of the parameters before scaling them.

\subsection{Neural network model}
\label{subsec-nn}

\begin{figure}
\centering
\includegraphics[width=\textwidth]{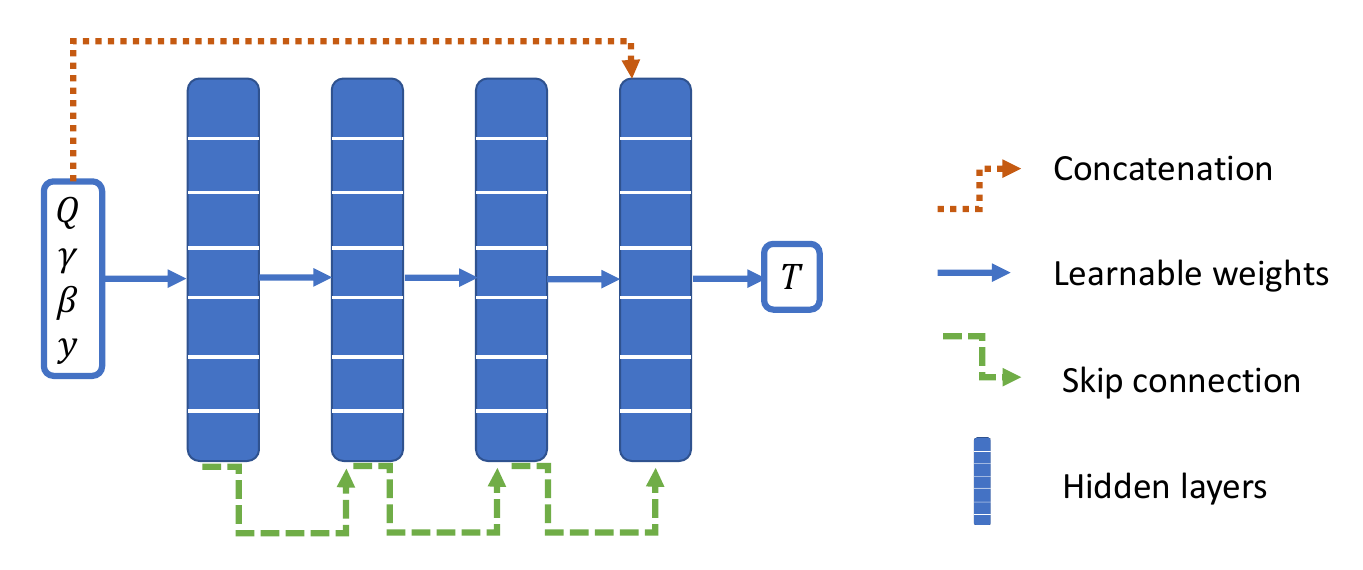}
\caption{We use a feedforward neural network to predict the temperature at a given height $y$ as a function of the simulation parameters: internal heating ($Q$), viscosity contrast due to temperature ($\gamma$), and viscosity contrast due to pressure ($\beta$). This pointwise prediction, instead of providing the complete temperature profile at once, helps prevent oscillations in the output of the network. The network has skip connections to regularize the loss landscape. We also condition the last hidden layer with the input vector by concatenation, as this tends to improve the accuracy of the predictions. Although not shown in the figure, we apply $SELU()$ activation to each hidden layer, i.e., each layer except the input and the output.}
\label{fig-nn-schema}
\end{figure}

We use a typical feedforward neural network, such as the one used in \cite{agarwal2020}, but modify it for our purposes (see Fig. \ref{fig-nn-schema}):

\begin{itemize}

\item Unlike the NN in \cite{agarwal2020}, the network output is a scalar prediction of temperature at a given height ($T(y)$). Such coordinate-based networks are heavily used for more memory-efficient learning on unstructured representations such as point clouds and meshes \citep[e.g.,][]{deepsdf, hines2023}. This also helps avoid wiggles in the prediction, which we see when the profile is high-dimensional (more than $\approx60$ points, as empirically observed in \cite{agarwal2020}). This also allows the end-user to directly train and evaluate on arbitrary depth profiles that are no longer tied to the underlying discretization. 

\item We train on mini-batches of only 32 points. Therefore, we find it beneficial to sample comparatively more points in the upper ($100$ times between from $15$th point from top) and lower ($50$ times from the $10$th point from bottom) thermal boundary layers of the temperature profile as these would otherwise be underrepresented in a batch as most of the mantle tends to be isothermal due to convective heat transfer. 
\item Skip connections are introduced from each hidden layer to every following hidden layer by addition before applying the activation function $SELU()$.
\item We inject the input to the NN into the last hidden layer of the network by concatenation, as it has shown to be useful for improving accuracy \citep[e.g.,][]{deepsdf}. 
\item Finally, we apply a correction at the boundaries of the domain. As NNs are approximators optimized by an iterative process, their predictions naturally tend to be inexact and have a hard time satisfying boundary conditions. In other words, the boundary conditions tend to be no more or less accurate than any other point in the domain. In the future, it could be worth exploring some non-trivial approaches for exact enforcing of the boundary conditions \citep[e.g.,][]{epinn,kumardistance}, but here we simply overwrite the NN predictions at the top with $0$ and at the bottom with $1$. However, this also means that the gradient of the boundary layer can become inconsistent, especially if the grid resolution is changed. To ensure that at least the heat fluxes are consistent with respect to the discretization, we overwrite the points within a certain distance of the top $(0.04)$ and bottom $(0.985)$ with the same gradient between the first and last point. 

\end{itemize}

We train the NNs in PyTorch \citep{pytorch2} using an Adam optimizer \citep{adam}. To keep over-fitting in check, especially for such a small dataset, we use a batch size of $32$, a weight decay value of $0.0005$ with the Adam optimizer and save the weights of the NN after an epoch only if the loss on the cv set has decreased. We train for $140$ epochs and decrease the initial learning rate of $0.001$ by a factor of $0.5$ every $20$ epochs. On a Tesla V100 GPU, training takes up to a couple of minutes. In contrast to \citet{agarwal2020}, the combination of small mini-batches, $SELU()$ activation function \citep{selu} instead of $tanh()$, pointwise formulation (using $y$ as an input) and a small number of simulation profiles (128) compared to $10,000$ simulations $\times 100$ time-steps makes for a significantly faster training time. We tested NN architectures by varying the number of hidden layers from $2$ to $6$ and the number of hidden units per layer between $32$ and $256$, and found $4$ or $5$ hidden layers with $128$ or $256$ units to perform the best. Here, we present the results of a NN trained with $128$ units with $5$ hidden layers as this achieved the lowest error on the cv set. 

\subsection{Regression baselines}
\label{subsec-baseline}

For a fair comparison, we include the following baseline algorithms for predicting the temperature profile using scikit-learn \citep{scikit}: 

\begin{itemize}
    \item Linear regression: we test linear regression to predict the temperature profiles as a function of the normalized simulation parameters. 
    \item Kernel ridge regression: it is a standard workhorse for regression problems and has the advantage of being able to approximate non-linear functions while being robust to outliers due to regularization. We use a radial basis function kernel with $\alpha=0.1$, which controls the regularization strength, and leave the default value of the multiplier of the radial basis function.
    \item Nearest neighbor: as a simple baseline, we checked how far off the temperature profile would be for a set of parameters if we simply picked the profile at the nearest neighbor in terms of the simulation parameters in our dataset. The nearest neighbor was determined using KDTree, which uses the Minkowski distance in an $n$-dimensional parameter space.
    \item Nearest neighbor interpolation: as a slightly more advanced version of nearest neighbor, we also baseline by taking (1) the temperature profiles of the three nearest neighbors in terms of simulation parameters and (2) by using inverse distance weighting to sum the three profiles resulting in an interpolated result. If the set of parameters being evaluated matches a set of parameters in the ``training'' dataset, we do not perform the averaging and simply take the exact output from the dataset.
\end{itemize}

\subsection{Derivation of basic scaling laws for convection}
\label{subsec-lid}

As an example of downstream applications of the kind of simulations we aim to accelerate in this study, we derive some basic scaling laws for mixed heating with temperature- and pressure-dependent viscosity. There are three main steps for deriving the scaling laws. 

\begin{figure}
\centering
\includegraphics[width=0.8\textwidth]{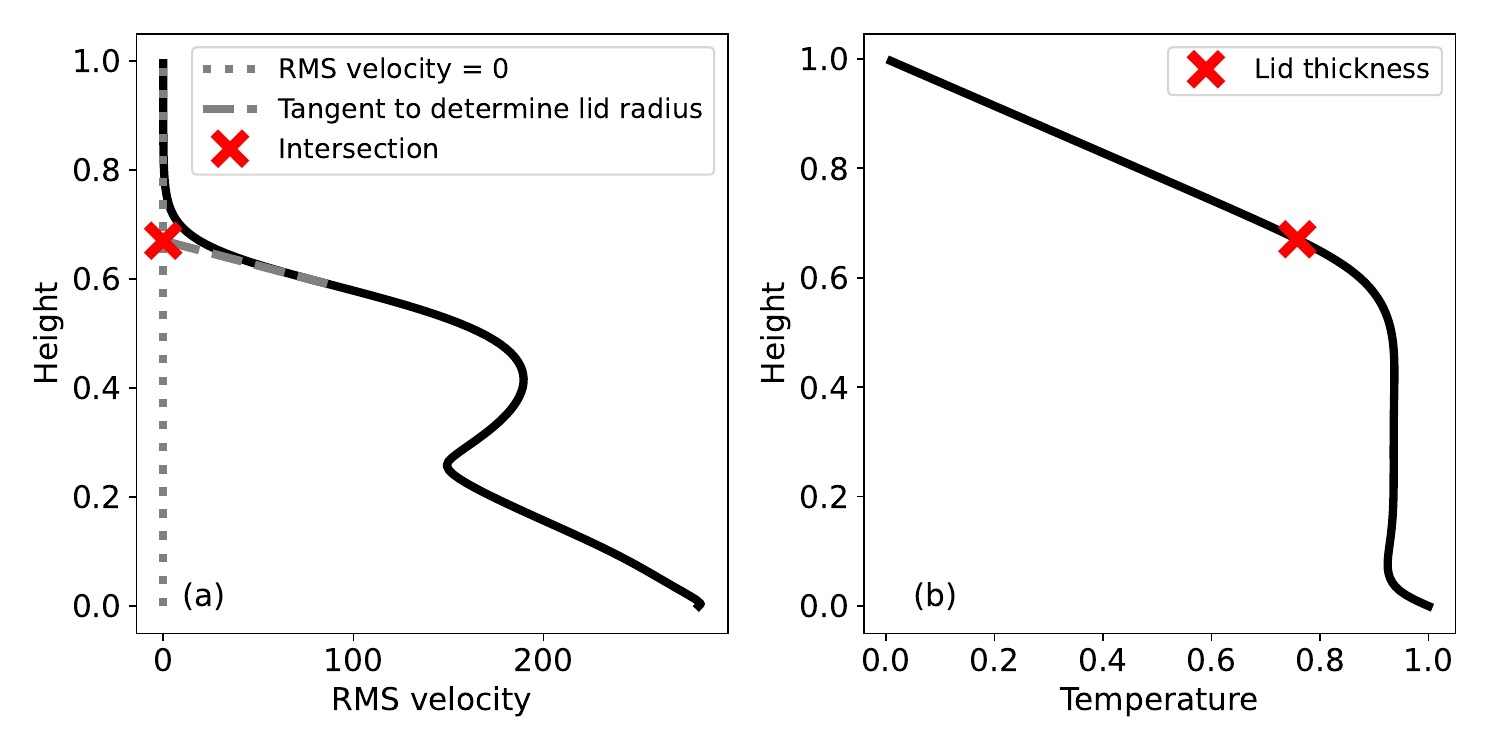}
\caption{The ``tangent method'' is used for determining the thickness of the lid. a) A tangent is drawn from the point of maximum gradient of the horizontally averaged profile of the RMS velocity to where it equals zero. This is taken as the lid thickness from top or the height of the lid in terms of the y-coordinate. b) The temperature contrast between this point and the maximum temperature (i.e. $T=1$ at the bottom of the domain) is used to determine an effective Rayleigh number of the system with respect to which scaling laws are derived.}
\label{fig-lid-thickness}
\end{figure}

First, we determine the thickness of the stagnant lid using the ``tangent method'' \citep[e.g.,][]{reese1999b} as shown in Fig. \ref{fig-lid-thickness}. The point at which the tangent from the point of maximum gradient of the horizontally-averaged root mean squared (RMS) velocity profile intersects the $y$-axis, i.e. where RMS velocity equals zero, is taken as the height of the lid. 

Second, the difference between the temperature at the base of this lid and the maximum temperature is used to calculate an effective Rayleigh number ($Ra_{\rm eff}$):
\begin{linenomath*}
\begin{equation}
    Ra_{\rm eff}=Ra \frac{y_{\rm lid}^3 (T_{\rm max} - T_{\rm lid})}{\eta_{\rm eff}},
    \label{eq:Ra-eff}
\end{equation}
\end{linenomath*}
where, $Ra$ is the reference Rayleigh number defined in eq. \eqref{eq:Ra}, $y_{\rm lid}$ is the height of the lid, $T_{\rm max}$ is the maximum temperature of the profile, $T_{\rm lid}$ is the temperature at the base of the lid, and $\eta_{\rm eff}$ is calculated as the harmonic mean of the viscosity below the lid.

Third, the simulations are used to derive a Nusselt-Rayleigh scaling. As this is not the primary goal of this study, we avoid a comprehensive survey of and comparison to the vast literature on scaling laws and instead use a basic Nusselt-Rayleigh scaling of the form proposed by \citet{moore2008} that considers an additive term for internal heat production:
\begin{linenomath*}
\begin{equation}
    Nu= a \ Ra_{\rm eff}^b + c \ Q,
    \label{eq:Ra-scaling}
\end{equation}
\end{linenomath*}
where $Nu$ is the Nusselt number at either the top or the bottom of the domain, and $a$, $b$ and $c$ are coefficients that are determined from simulations accelerated by the NN predictions with the help of the curve fitting function in Scipy \citep{scipy}, which accepts arbitrary expressions such as the powerlaw function defined in eq. \eqref{eq:Ra-scaling}.  

\section{Results and Discussion}

\subsection{Neural network predictions vs. regression baselines}

\label{subsec-prediction}
\begin{table}
    \centering
    \begin{tabular}{cccc}
        Prediction algorithm    & MAE train & MAE cv & MAE test \\
        \hline 
        Linear regression       & 0.0385           & 0.0388 & 0.0676   \\
        Kernel ridge regression & 0.0148           & 0.0147 & 0.0371   \\
        Nearest neighbor        & \textit{0.0000}  & 0.0282 & 0.0495   \\
        Interpolation           & \textit{0.0000}  & 0.0230 & 0.0448  \\
        Neural network          & 0.0048           & \textit{0.0053} & \textit{0.0133}
    \end{tabular}
    \caption{Mean absolute error (MAE) of all the prediction algorithms on the train, validation (cv) and test sets. In the test set, at least one of the three simulation parameters is extrapolated. For completeness, the MAE for the training dataset is also presented for nearest neighbor and nearest neighbors interpolation although it is obvious that when the simulation with the same parameters in the dataset is present, the error will be exactly zero.}
    \label{tab-mae}
\end{table}

In this section, we present the results of predictions from an NN (see Sec. \ref{subsec-nn} for details) with $5$ hidden layers containing $128$ units each and compare these to the predictions of the regression baselines outlined in Sec. \ref{subsec-baseline}. Table \ref{tab-mae} shows that the NN achieves the lowest mean absolute error (MAE) of all the algorithms on the cv and test sets. Of course, in the case of nearest neighbor and nearest neighbors interpolation, the predictions on the training set are redundant and the MAE is exactly zero. The NN seems to generalize not only for parameters within the bounds of the training dataset (cv), but also generalizes better than the other algorithms when extrapolating slightly (test).  

\begin{figure}
\centering
\includegraphics[width=\textwidth]{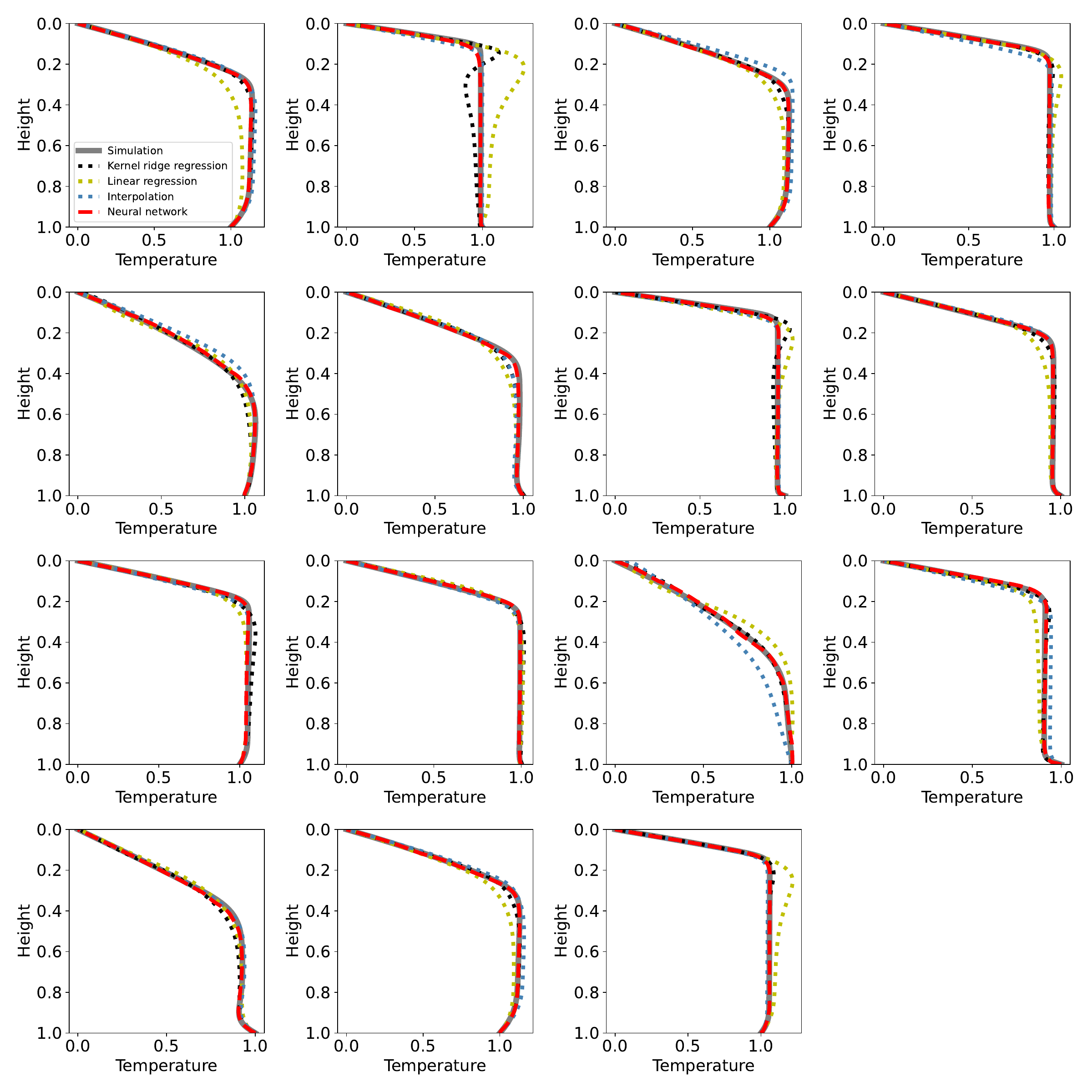}
\caption{Comparison of neural network predictions and those of the baseline algorithms against ground truth temperature profiles for the validation (cv) set.}
\label{fig-profs-cv}
\end{figure}

\begin{figure}
\centering
\includegraphics[width=\textwidth]{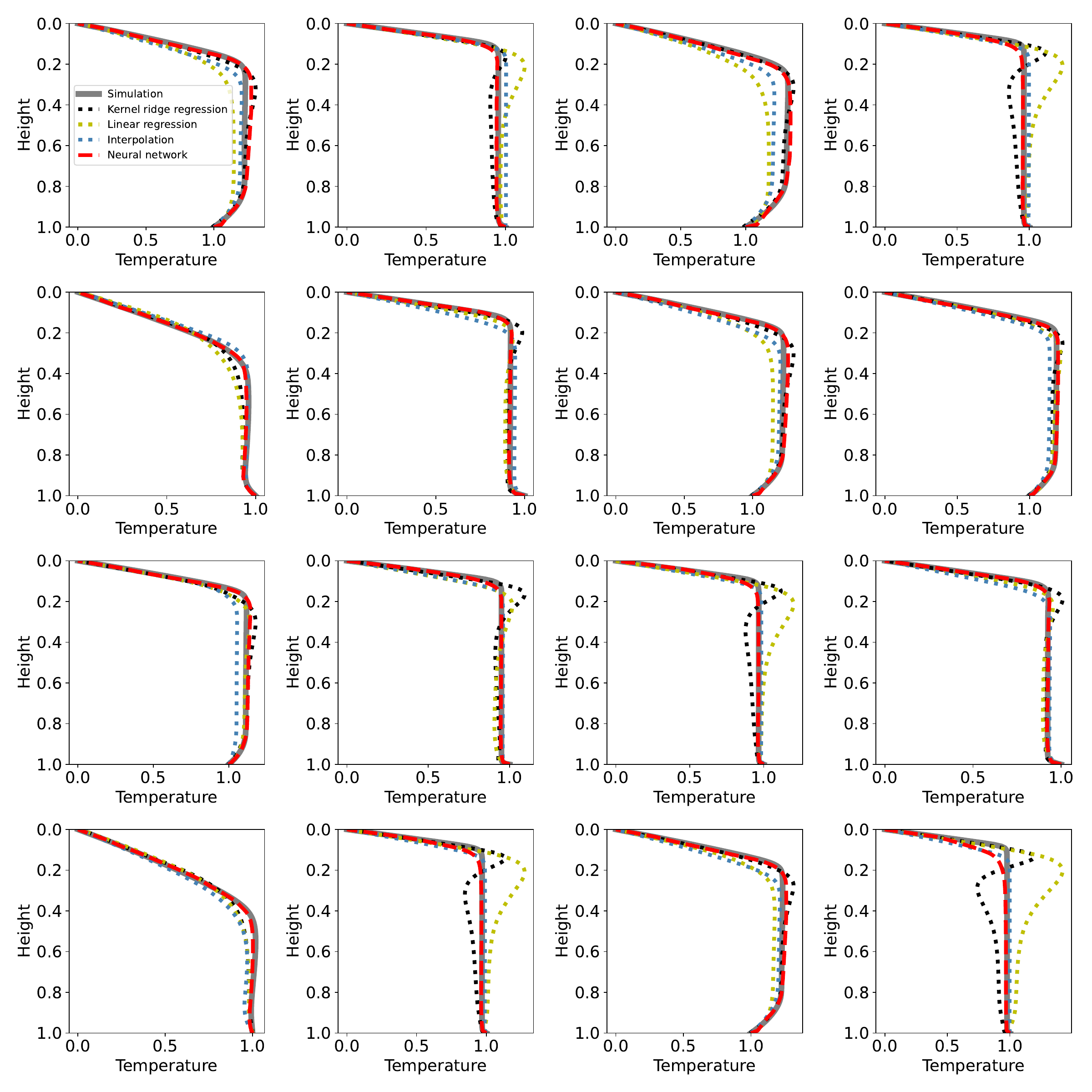}
\caption{Comparison of neural network predictions and those of the baseline algorithms against ground truth temperature profiles for the test set.}
\label{fig-profs-test}
\end{figure}

We plot the predictions of the NN and of the regression baselines against the ground truth in Fig. \ref{fig-profs-cv} for the cv set and in Fig. \ref{fig-profs-test} for the test set. In both cases we leave out the nearest neighbor prediction and take the interpolated profile based on the nearest neighbors instead, which is based on the three closest neighbors and is therefore slightly more accurate. Visually, we can observe that the NN predictions seem to match the ground truth most consistently. Linear regression and kernel ridge regression sometimes introduce larger wiggles at the base of the lid. The nearest neighbors interpolation produces visually reasonable results, but ultimately suffers from low accuracy.

\begin{figure}
\centerline{\includegraphics[width=1.4\textwidth]{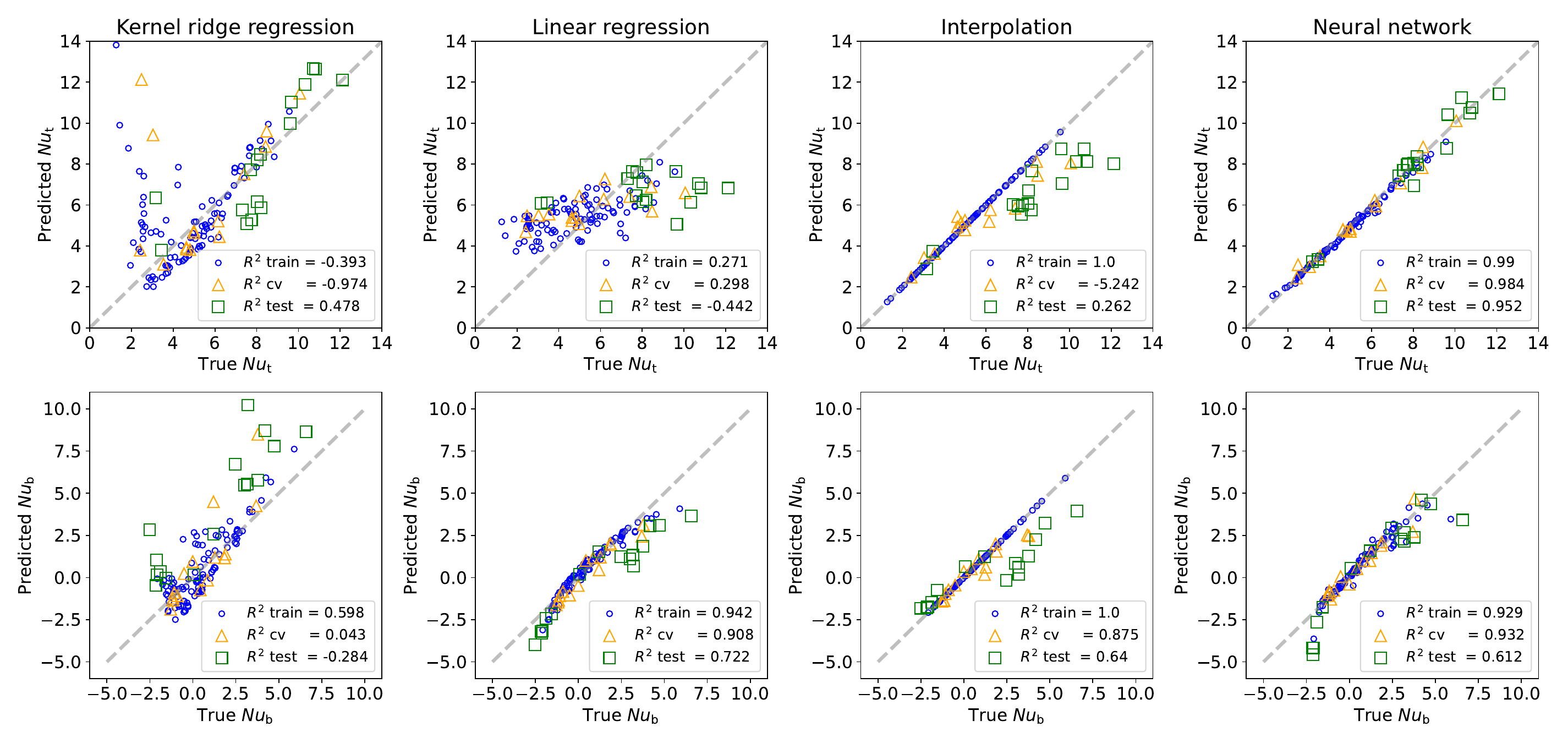}}
\caption{Comparison of neural network predictions and those of the baseline algorithms against the true non-dimensionalized heat flux at the top $Nu_{\rm t}$ (top row) and at the bottom $Nu_{\rm b}$ (bottom row).}
\label{fig-pred-nu}
\end{figure}

We further use the temperature profiles to calculate the top and bottom Nusselt numbers, i.e. the non-dimensional heat fluxes computed at the upper and lower domain boundary. As shown in Fig. \ref{fig-pred-nu}, even though the NN predictions are not accurate enough to derive scaling laws, they outperform the baselines. The NN accuracy diminishes when extrapolating and at least at the bottom, it fares slightly worse than linear regression and interpolation in the test set. This is because the thermal gradients at the bottom are steeper than in the top thermal boundary layer and are therefore more sensitive to slightly perturbed (inaccurate) predictions.

\subsection{Accelerating convergence of numerical simulations}
\label{subsec-speedup}
We demonstrate the impact of different initial conditions on the time it takes to reach a statistical steady-state. For this, we consider three qualitatively different simulations by varying the three parameters $Q$, $\beta$ and $\gamma$ as follows: 
\begin{itemize}
    \item Case 1 (weak convection):   $Q=5$, $\gamma=10^8$, $\beta=50$
    \item Case 2 (medium convection): $Q=7.5$, $\gamma=10^9$, $\beta=25$
    \item Case 3 (strong convection): $Q=10$, $\gamma=10^{10}$, $\beta=1$
\end{itemize}
For each of these three simulations, we then start from the following initial temperature profiles: (1) hot, with $T=1$ everywhere except at the top boundary; (2) cold, with $T=0$ everywhere (except at the bottom boundary); (3) linear, with $T$ increasing linearly from the top, where $T=0$, to the bottom, where $T=1$, as in a conductive profile; (4) perfect, corresponding to the last time step of a finished numerical simulation; and (5) NN prediction. For the weak and medium convection cases, we show the perfect restart for reference to get a sense of how the simulation would evolve if our temperature profile predictions had no error. For the strong convection case, the simulation could not reach a statistical steady-state after a few weeks and therefore, no perfect restart is available. 

\begin{figure}
\centerline{\includegraphics[width=1.4\textwidth]{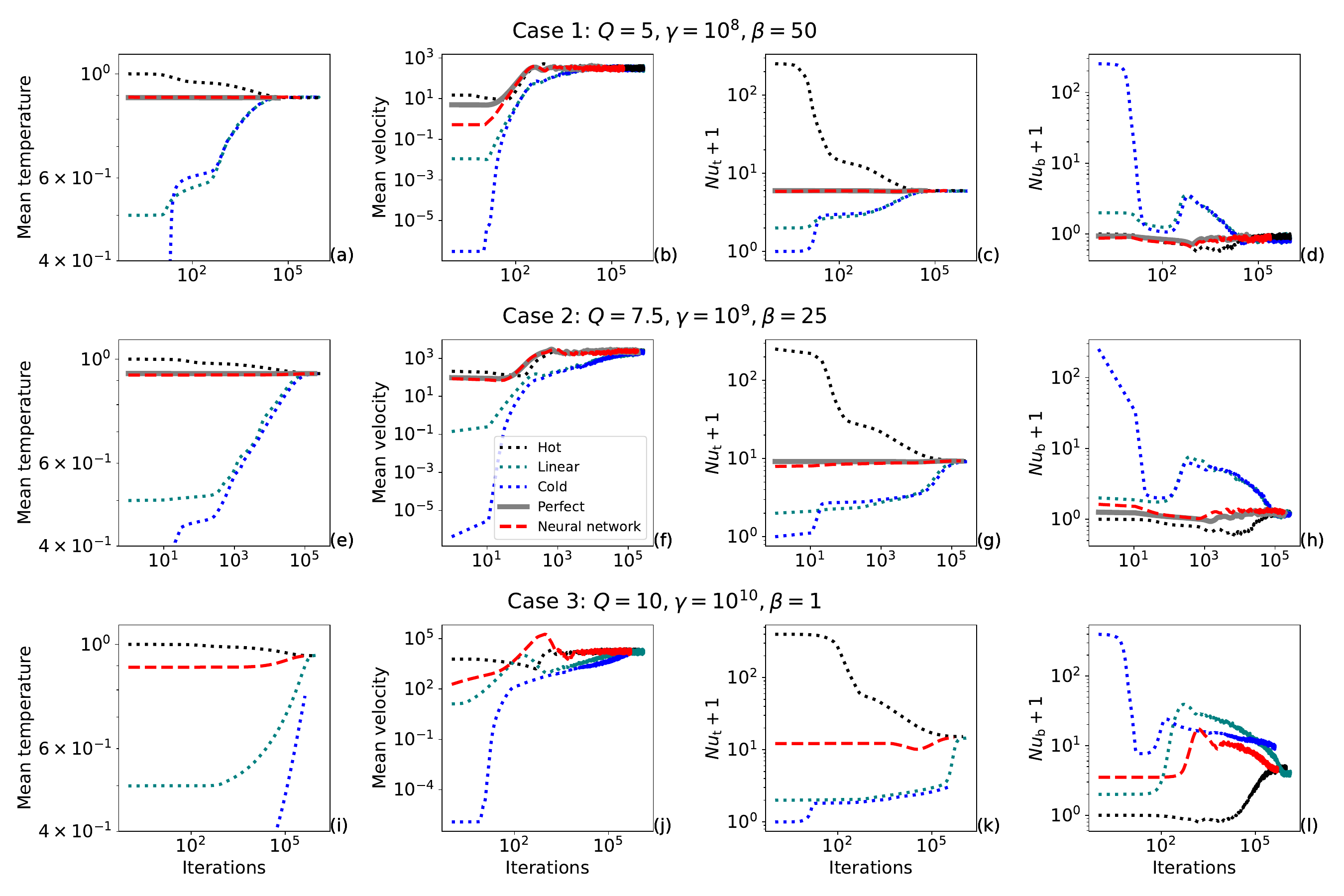}}
\caption{Time-series of three different simulations started with different initializations: (1) hot, (2) cold, (3) linear, (4) perfect, i.e. the exact profile from a finished simulation for reference, and (5) NN prediction. To enhance the visualization, we use the $\log_{10}$-$\log_{10}$ scale for plotting and therefore add one to the heat fluxes to ensure that the $\log_{10}$ is defined. For the strongly convecting case (row 3), a definitive steady-state could not be achieved within the time-frame of this study. However, it seems that the hot, linear and neural network initializations all begin to converge to the same value after a similar number of iterations. It is interesting to note that the internal heating ($Q$) and viscosity contrast due to temperature ($\gamma$) are beyond the range of the training data of the neural network and therefore, the imperfect prediction fails to offer any speedup compared to the hot and linear temperature profiles.}
\label{fig-sims}
\end{figure}

\begin{table}
    \centering
    \begin{tabular}{cccc}
        Initialization    & Case 1  & Case 2 \\
        \hline 
        Cold            & 45,909  &  212,906   \\
        Hot             & 47,948   &  115,950  \\
        Linear          & 101,802   &  194,898   \\
        Neural network  & \textit{17,624}  &  \textit{68,058} \\
        \hline
         Perfect         & 5,593   &  21,368  \\
        \hline
    \end{tabular}
    \caption{Iterations needed for mean temperature, RMS velocity and heat fluxes at top and bottom to reach within $1\%$ of the final value averaged over the last $10\%$ of total iterations. Initializing the simulations with the profiles predicted by the NN (last line) results in the minimum number of iterations needed to reach a statistical steady state. For reference, we provide the iterations needed for the perfect restart case as an upper bound of the speedup that can be expected. For case 2, this upper bound of speedup from a perfect profile would be $5.4$.}
    \label{tab-sim}
\end{table}

Fig. \ref{fig-sims} shows the mean temperature, mean velocity, Nusselt number at the top and Nusselt number at the bottom for all the simulations and initializations as a function of the number of time steps of the numerical solver (indicated as Iterations in the figure). For the weak and medium convection cases (Case 1 and 2, respectively), the NN-initialized simulation (dashed red line) follows the simulation with the``perfect'' temperature profile initialization (solid grey line) very closely. The other initializations are naturally very far off at the beginning of the simulation and typically take on the order of $10^4$ iterations to reach a statistical steady-state. In Table \ref{tab-sim}, we list the number of solver iterations needed to reach within one percent of the final mean temperature, RMS velocity and heat fluxes at top and bottom averaged over the last $10\%$ of total iterations. For reference, we also provide the iterations needed for the perfect restart case. This serves as an upper bound of the speedup that can be achieved.

\begin{figure}
\centerline{
\includegraphics[width=1.4\textwidth]{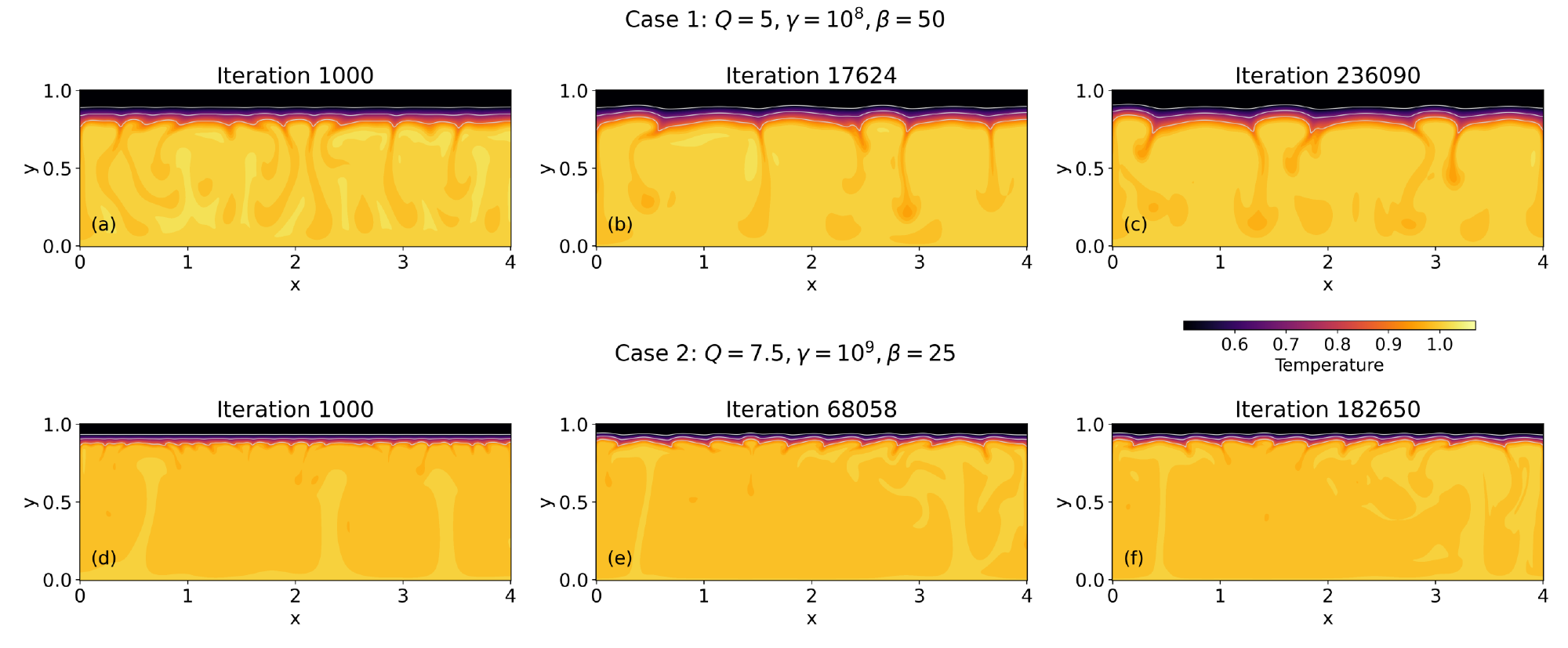}}
\caption{Snapshots of the temperature fields for the weak (a-c) and medium convection (d-f) cases at $1000$ iterations (a,d), at the iteration identified as the statistical steady-state from the mean quantities (see Fig. \ref{fig-sims} and Table \ref{tab-sim}) (b,e), and the last iteration available for the simulation (c,f).}
\label{fig-snapshots}
\end{figure}

Deciding when a simulation has reached a statistical steady-state is not always straightforward. For example, if the goal is to derive a type of Nusselt-Rayleigh scaling law, then the stability of the mean quantities such as the Nusselt numbers plays a key role. However, if one wishes to retrieve the two-dimensional temperature field, it might be also important to consider if this field has stabilized or not. Fig. \ref{fig-snapshots} shows some snapshots for Cases 1 and 2, when started from the NN initialization: $1000$ iterations after the restart (a, d), at the iteration where the statistical steady-state is reached (b, e), and at the final iteration available (c, f). Even though the top Nusselt number stabilizes for the first simulation (Fig. \ref{fig-sims}c) almost immediately at the start, the horizontal distribution of the stagnant lid takes longer to develop (Fig. \ref{fig-snapshots}a-c). Hence, deciding when steady-state has been reached is a decision best left for the informed user to make based on the purpose of the study. 

The NN outperforms the other initializations for the weak convection case (Case 1) and the medium convection case (Case 2). The starting heat flux for Case 2 is slightly off and takes quite a few iterations to reach the final value, hinting that even a slight error in the initial temperature profile can have a large effect on the speedup factor. In the future, it would be worth improving the accuracy of the NN predictions, as better predictions are likely to offer significantly larger speedups (as seen from perfect restart in Table \ref{tab-sim}).

We now return to Case 3, the strongly convecting case which has run for several weeks. Although, the perfect restart is not available, Fig. \ref{fig-sims} seems to indicate that the hot, linear and NN restarts are converging to the same point. Again, this seems to reinforce the intuition that an initial condition that is too far off from the final state is not any better than some other typically used initializations. At the same time, it is important to note that this prediction is especially challenging for the NN, because here we are extrapolating in two parameters: $Q$ and $\gamma$. In light of this, it is interesting that the NN-initialized simulation is not slower than the other initializations in reaching what seems to be the final steady. In other words, within reasonable departures from the parameter space, using the NN initialization at least does not harm the speed of convergence. 

\subsection{Scaling laws for convection with variable viscosity and internal heating}
\label{subsec-scalinglaw}
As an example of an application of simulations accelerated by NN predictions, we derive some basic scaling laws for three cases, where we vary the maximum viscosity contrast due to temperature ($\gamma$), and thereby the effective Rayleigh number: 
\begin{enumerate}
    \item no internal heating and no pressure dependence of viscosity: $Q=0$ and $\beta=1$
    \item internal heating and no pressure dependence of viscosity: $Q=2$ and $\beta=1$
    \item internal heating and pressure dependence of viscosity: $Q=1$ and $\beta=10$
\end{enumerate} 
To generate the data needed to fit the coefficients of eq. \eqref{eq:Ra-scaling}, we run eight simulations for each of the cases listed above and vary $\gamma$  between $10^6$ and $10^9$ in $\log_{10}$ space. We run these simulations starting from the temperature profile predicted by the trained NN. For $Q=1$, $\beta=10$ and $\gamma=10^6$, the simulation was purely conductive and hence discarded.

\begin{figure}
\centering
\includegraphics[width=\textwidth]{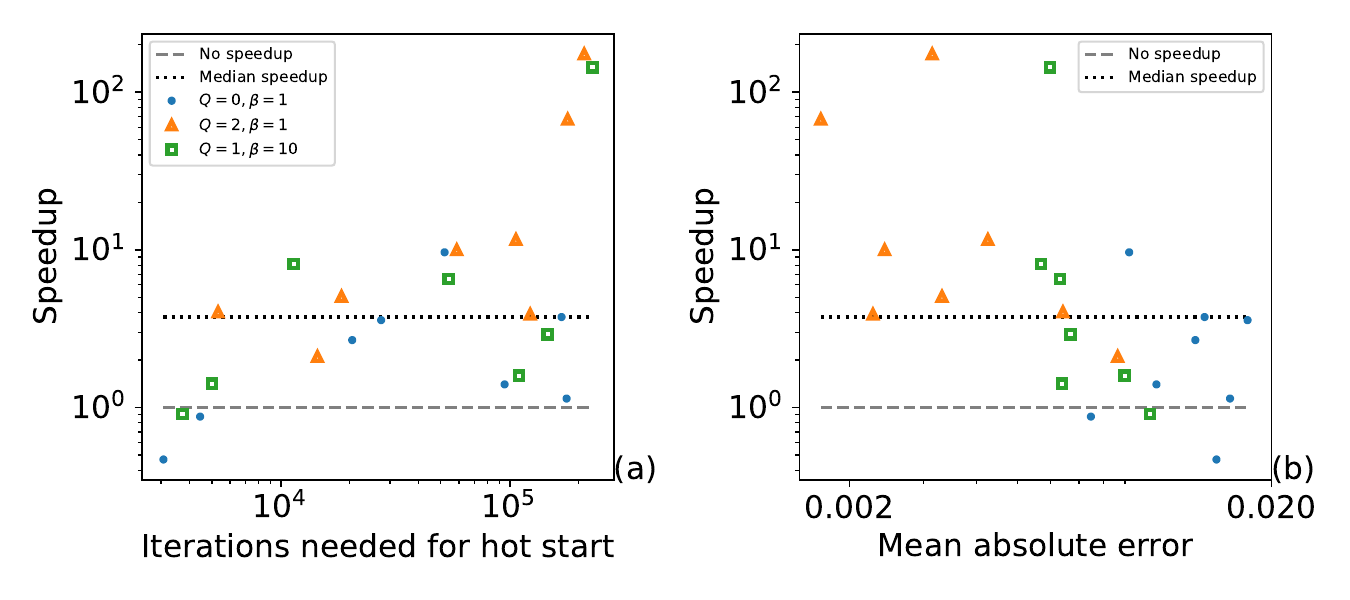}
\caption{For the simulations used to derive scaling laws, two types of initializations are used: NN and hot. The number of iterations needed by the simulations with hot starts are divided by those initialized with the NN predictions to arrive at the speedup factor. The speedup factor is plotted as a function of (a) the iterations needed by the hot starts and (b) the mean absolute error between the initial condition, i.e. the NN prediction and the final temperature profile of the statistically steady-state system (obtained by averaging the temperature profiles over the last $1\%$ of the iterations). Despite the scatter, there is a negative correlation with respect to the prediction error in the profiles, suggesting that the more accurate the initial condition of the system is, the faster it converges. The median value of speedup here is $3.75$.}
\label{fig-speedup}
\end{figure}

To enrich the investigation in Sec. \ref{subsec-speedup} on the speedup offered by our NN, we run the same set of $24$ simulations with the so-called hot start and compare the two initializations in terms of the iterations needed to reach within $1\%$ of the final value of the mean temperature, RMS velocity and Nusselt numbers at the top and bottom. Fig. \ref{fig-speedup} shows the speedup defined as the number of iterations needed by the simulations with hot start divided by the number of iterations needed by the NN-initialized simulations. With respect to $Q$, $\gamma$ and $\beta$, no clear pattern is evident in the speedup factor, which is scattered across a few orders of magnitude: from being even slower ($0.5$ times) to being significantly faster ($174$ times). However, in Fig. \ref{fig-speedup}(a), we observe a positive correlation between the speedup factor and the number of iterations needed by the hot start to reach the steady-state, indicating that the more computationally challenging simulations often benefit more from this method. In Fig. \ref{fig-speedup}(b), we observe a negative correlation between the speedup factor and the mean absolute error between the final temperature profile of the system at statistical steady-state and the starting profile, reinforcing our intuition from Fig. \ref{fig-sims}. For the simulations here, the NN accelerates the attainment of a statistical steady-state in $86\%$ of the cases. Including simulations without any observed speedup, the median factor is $3.75$. 

\begin{figure}
\centering
\includegraphics[width=\textwidth]{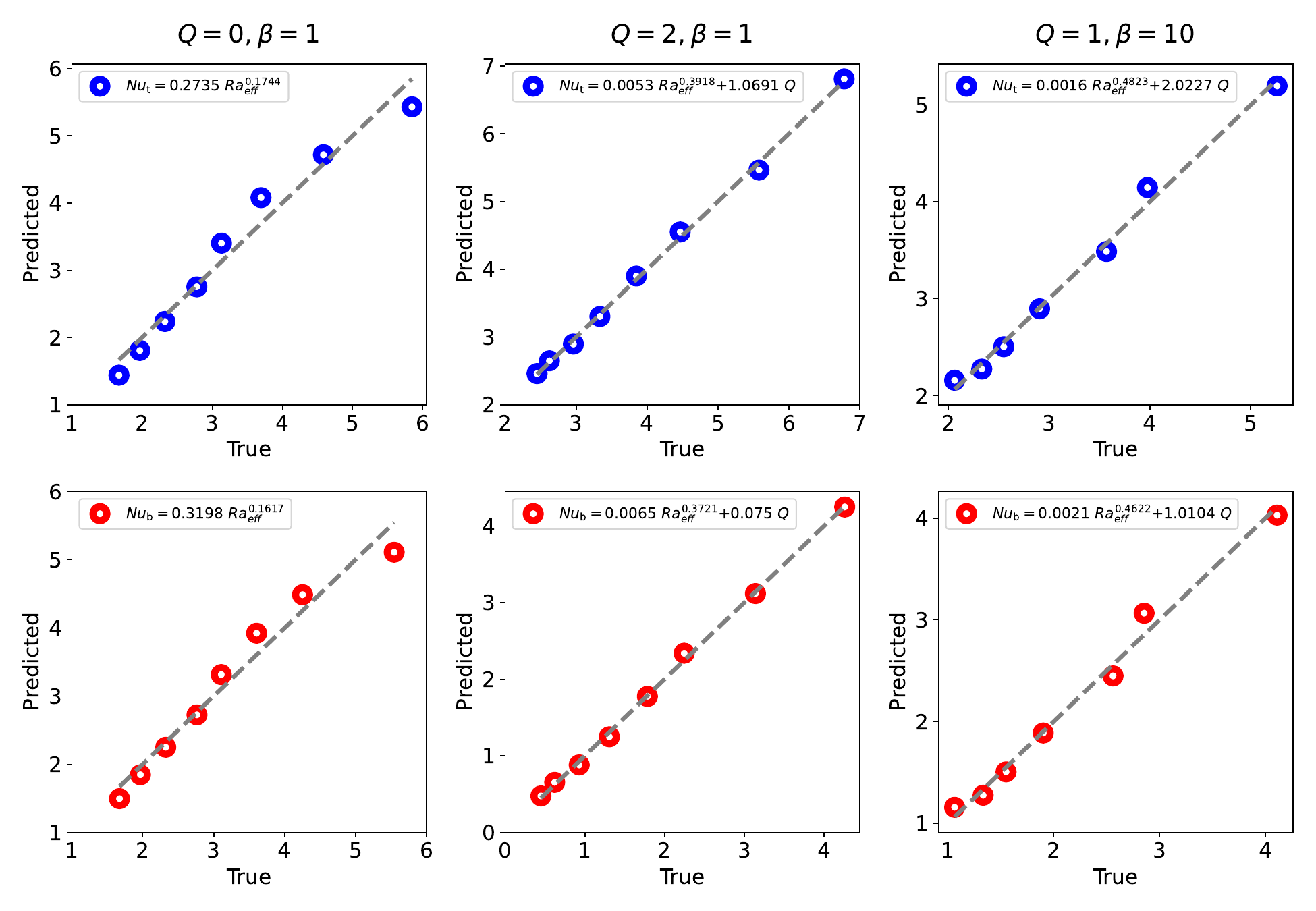}
\caption{As an example of typical applications of such steady-state simulations, we derive basic scaling laws for three cases: no internal heating and no pressure dependence of the viscosity (left column), internal heating with $Q=2$ and no pressure dependence (middle column), and internal heating with $Q=1$ and pressure-dependent viscosity with $\beta=10$ (right column).}
\label{fig-scaling-laws}
\end{figure}

Using these simulations, we derive scaling laws for each of the three scenarios described at the beginning of this subsection. We then use these scaling laws to predict the Nusselt numbers at the top ($Nu_t$) and bottom ($Nu_b$), and plot them against the true values from the simulations in Fig. \ref{fig-scaling-laws}. Despite some scatter, the scaling laws are able to predict the Nusselt numbers at top and bottom well. Our approach thus paves the way to efficiently generate large datasets of multi-parameter simulations that can be used to predict generalized scaling laws for complex flows with variable viscosity and heating mode, relevant for planetary interior evolution modeling. 

\section{Discussion}
\label{sec-discussion}

While sophisticated machine learning methods have demonstrated remarkable capabilities in approximating highly non-linear forward maps, they often tend to be data-intensive. This is also true for mantle convection, where surrogate models must be a function of simulation parameters and thereby learn features on vastly different spatio-temporal scales. For example, \citet{agarwal2021b} were able to learn the two-dimensional evolution of temperature fields, but they needed more than $10,000$ simulations and did not predict other system variables such as flow velocities. The predictions also had errors with respect to the numerical simulations and the two-step learning procedure (dimensionality reduction with autoencoders and time-stepping with LSTMs) took almost a week to train on a single GPU. 

This study provides an alternative approach where only $97$ simulations were used to learn the profiles with a training time of approximately $2$ minutes. No High Performance Computing resources were used. Also, there is no prediction error as the method is used to initialize a numerical solver. At the same time, this study demonstrates a median speedup of $3.75$ times, whereas \citet{agarwal2021b} demonstrated a speedup on the order of $10,000$. Although, a one-to-one comparison of the two approaches is non-trivial due to their different goals, setup of the simulations and learning tasks, it seems to be the case that the more computational effort is put into learning (data generation plus optimizing the neural network), the greater the potential speedup. We show how one can already start reaping efficiency gains from machine learning models trained on small datasets.

While methods for advancing the full system states in time remain an interesting alternative avenue of research -- even for simulations aimed at reaching a statistical steady-state -- the approach presented in this paper can also benefit from some improvements:

\begin{itemize}
    \item Since higher accuracy in the prediction of temperature profiles generally led to a higher speedup, it would be worth improving the accuracy of the NN. One way to do so would be to add more data. Interestingly, these new simulations could already be initialized with the predictions of the current NN and hence potentially faster to generate. In the future one can even consider an incremental dataset generation approach, where the NN trained on the previous chunks of data is used to accelerate the next batch. This approach has some interesting parallels with simulation-based-inference methods where evaluations on limited data guide how later simulations are generated \citep[e.g.,][]{PapamakariosMurray2016}.

    \item It would be remiss of us to generate larger training datasets without also extending the setup of the physical simulations. It would be interesting for mantle convection studies to incorporate more realistic geometries such as the spherical annulus \citep{herlund2008,fleury2024}, and to vary the ratio of core to surface radius in order to make the results readily relevant for different planetary bodies.
    
    \item Our initial attempts to learn a two-dimensional temperature field using the pointwise neural network were not successful, but it could be worth exploring other learning strategies to overcome the uncertainties introduced by the fluctuating components of the temperature fields. One example could be to learn a set of phase-shift features whose sole job is to help distinguish one snapshot of the temperature field in the statistical steady-state from another where only some plumes and downwellings have shifted. One could then produce any reasonable realization of this set of features as an initial condition and hopefully then attain the steady-state almost immediately with the solver. Alternatively, more recent advances in generative modeling for fluid flows could also be a promising area of research \citep[e.g.,][]{lienen2024zero, saydemir2024unfolding}.
    
    \item With this extended physical setup and possibly an even greater speedup from the improvements mentioned above, one would then be in the position to generate a large dataset and derive generalized scaling laws in a spherical annulus geometry that incorporate mixed heating as well as pressure- and temperature-dependence of viscosity. 
\end{itemize}

\section{Conclusions}\label{sec-conclusions}

We presented a method to accelerate the computationally-intensive determination of the statistical steady state of mantle convection simulations. Using a neural network (NN), one can learn the final horizontally-averaged temperature profile of the system. For unseen simulations, this prediction serves as an efficient initial condition for the solver which attains a statistical steady-state significantly faster than with other typically used initial conditions. While the NN prediction is already quite accurate in most cases, running the solver enables one to obtain higher order statistics of the system such as the two-dimensional temperature fields and that too at numerical accuracy, i.e. without any prediction error as the method initializes a numerical solver.  

Even in the limit of the few simulations ($97$) used to train the NN, we found that it outperformed other regression baselines such as linear regression, kernel ridge regression and nearest neighbor interpolation. This was not only true when interpolating in the parameter space, but also when slightly extrapolating in at least one parameter. Nevertheless, the results indicated that it was not any faster to start from an NN prediction if the parameters were outside the training range and had therefore a higher error. This is because any error in the starting temperature profile takes several solver iterations to be corrected, thereby negating the advantage of such an initialization. Still, it is encouraging that for the $23$ simulations we ran to derive some basic scaling laws, we observed a speedup by our method in $86\%$ of the cases. Including cases where no speedup was observed, the simulations initialized with NN predictions reached a steady-state $3.75$ times faster (median value) than those initialized with hot profiles.

From a machine learning for fluid dynamics perspective, this study falls on the less resource-intensive side, requiring fewer than $100$ simulations with a training time of $2$ minutes for the NN. While more sophisticated learning approaches can offer greater speedups, we have shown here that, when combined with domain expertise, machine learning can deliver actual benefits and can do so by speeding up traditional methods such as computational fluid dynamics instead of competing with them. We would like to extend the physical setup of the simulations to a spherical annulus and also vary the radii of the core to the surface as an additional parameter. We further propose an incremental generation of a larger dataset, where the NN trained on the finished simulations are used to initialize the new ones. With these simulations, one can expect an even more accurate NN, allowing derivation of general scaling laws that simultaneously incorporate internal heating and temperature- and pressure-dependence of viscosity.

Regardless of where one falls on the spectrum of investment vs. efficiency gains, machine learning promises to significantly accelerate mantle convection simulations and help improve our understanding of planetary interiors.

\section{Author Contributions}
We list the author contributions following the CRediT taxonomy. \\
\textit{Conceptualization}: CH, SA, NT, DG, AB; \\
\textit{Methodology}: SA, CH, NT, DG, AB ; \\
\textit{Software}: SA, CH ; \\
\textit{Validation}: SA, CH, NT  ; \\
\textit{Formal analysis}: SA, CH ; \\
\textit{Investigation}: SA, CH, NT  ; \\
\textit{Resources}: NT, CH  ; \\
\textit{Data curation}: NT, CH, SA ; \\
\textit{Writing–Original Draft}: SA, NT, CH, AB ; \\
\textit{Writing–Review \& Editing}: DG, AB, CH, NT, SA ; \\
\textit{Visualization}: SA, NT; \\
\textit{Supervision}: NT, DG;  \\
\textit{Project administration}: NT, DG, AB; \\
\textit{Funding acquisition}: NT, DG, SA; \\

\section*{Open Research Section}

The trained model can be run to generate and download temperature profiles: \url{https://huggingface.co/spaces/agsiddhant/steadystate-mantle}. No installation is needed. The repository also contains the raw dataset used to train, validate and test the machine learning models as well as the pytorch routines used to train: \url{https://huggingface.co/spaces/agsiddhant/steadystate-mantle/tree/main/data}.

\acknowledgments
This work was part of the PLAGeS (Physics-based Learning Algorithms for Geophysical flow Simulations) project and funded by the German  Ministry of Education and Research BMBF (project numbers 16DKWN117A and 16DKWN117B).

%
%

\bibliography{Bibliography}

%
%
%
%
%

\end{document}